\documentclass[conference]{IEEEtran}
\IEEEoverridecommandlockouts
\usepackage{cite}
\usepackage{amsmath,amssymb,amsfonts}
\usepackage{algorithmic}
\usepackage{graphicx}
\usepackage{textcomp}
\usepackage{xcolor}
\usepackage{float}
\usepackage{amsthm}
\usepackage{graphicx}
\usepackage{epstopdf}
\usepackage{bm}
\usepackage{color}
\usepackage{multirow}
\usepackage{multicol}
\usepackage{soul}
\def\BibTeX{{\rm B\kern-.05em{\sc i\kern-.025em b}\kern-.08em
    T\kern-.1667em\lower.7ex\hbox{E}\kern-.125emX}}
\begin{document}

\title{Max-Min Fair Wireless-Powered Cell-Free Massive MIMO for Uncorrelated Rician Fading Channels
\thanks{This work was partially supported by ELLIIT and the Wallenberg AI, Autonomous Systems and Software Program (WASP) funded by the Knut and Alice Wallenberg Foundation.}
}

\author{\IEEEauthorblockN{\"Ozlem Tugfe Demir and Emil Bj\"ornson}
	\IEEEauthorblockA{{Department of Electrical Engineering (ISY), Link\"oping University, Sweden
		} \\
		{Email: \{ozlem.tugfe.demir, emil.bjornson\}@liu.se}
	}
	
}

\maketitle

\begin{abstract}
This paper considers cell-free massive multiple-input multiple-output systems where the multiple-antenna access points (APs) assist the single-antenna user equipments (UEs) by wireless power transfer. The UEs utilize the energy harvested in the downlink to transmit uplink pilot and information signals to the APs. We consider practical Rician fading with the line-of-sight components of the channels being phase-shifted in each coherence block. The uplink spectral efficiency (SE) is derived for this model and the max-min fairness problem is considered where the optimization variables are the AP and UE power control coefficients together with the large-scale fading decoding vectors. The objective is to maximize the minimum SE of the users under APs' and UEs' transmission power constraints. An alternating optimization algorithm is proposed for the solution of the highly-coupled non-convex problem. 
\end{abstract}

\begin{IEEEkeywords}
cell-free massive MIMO, max-min fair power control, wireless power transfer, spectral efficiency, Rician fading
\end{IEEEkeywords}
\vspace{-0.4cm}
\section{Introduction}
Massive MIMO (multiple-input multiple-output) has been extensively studied for the cellular systems due to its high spectral efficiency (SE) achieved by spatial multiplexing of many user equipments (UEs) on the same time-frequency resource  \cite{emil_book,erik_book,ozge_massive,massive_mimo_reality}. Now, it is one of the key technologies in 5G and commercial deployments began in 2018 \cite{massive_mimo_reality}. However, the largest improvements are achieved by UEs that are close to a base station, while path loss and inter-cell interference will still lead to large SE variations \cite{emil_multiple}. Recently, an alternative network infrastructure to the cellular systems is considered in \cite{nayebi,cell_free_vs_small_cell}, which uses the name \emph{cell-free massive MIMO} since a large number of access points (APs) is distributed over a large geographic area to serve all the UEs without any cell boundaries. As shown in \cite{cell_free_vs_small_cell, making_cell_free}, the cell-free massive MIMO performs better than co-located massive MIMO and small cell systems in  providing uniformly good service to all the UEs. 

Communication and positioning are the main use cases for radio frequency (RF) in current wireless technologies. While we are in the era of 5G for mobile communication, some immature technologies have potential to be integrated into future generation standards. Wireless power transfer (WPT) via RF signals is one of these technologies and there has been extensive research conducted in this area to charge mobile battery-powered devices via the ambient RF signals \cite{1g,bruno}. WPT would reduce the battery requirements of the mobile devices and provide more consistent and ubiquitous service to energy-hungry devices. In particular, future autonomous low-power networks and Internet of Things (IoT) are expected to benefit from this technology \cite{1g}. An interesting paradigm in WPT is the simultaneous wireless information and power transfer (SWIPT) which was also considered for cellular massive MIMO systems \cite{ps_downlink2, swipt_downlink, swipt_downlink2}. In these works, the UEs have either a power splitting or time switching circuit to utilize the downlink RF signals for both information reception and energy harvesting. In \cite{wet2,wet_heath}, a base station (BS) enables uplink pilot and data transfer by energy beamforming in the downlink. In this paper, we adopt this setup for the cell-free massive MIMO which has a high potential to improve wireless-powered communication due to the reduced distances between the transmitter and receiver terminals and the increased number of energy sources compared to co-located massive MIMO.

The works which exploit WPT in cell-free systems are rather limited. In \cite{wpt1}, total harvested energy throughout of the network is maximized together with the AP selection under transmission power constraints for each AP. This work assumes perfect channel state information and does not take into account the uplink communications. In \cite{swipt_cell_free}, SWIPT is considered in the context of cell-free massive MIMO where information and energy UEs are located separately. Similarly, \cite{wpt2} studied cell-free massive MIMO where the information UEs do not harvest energy and there is a single energy-harvesting UE that actively eavesdrops. To the best of authors' knowledge, this paper is the first that considers power control for max-min fairness based uplink SE by downlink energy beamforming in the cell-free massive MIMO. Max-min fairness is one of the most studied optimization criteria for cell-free massive MIMO systems due to the fact that it maximizes the minimum guaranteed SE to all the UEs, which is highly in accordance with the uniformly great service motto of cell-free systems. Furthermore, max-min fairness may be effective to reduce the traffic congestion mainly resulting from UEs in bad channel conditions, by increasing the \%95-likely SE of the whole network. 
The main contributions of this paper are:
\begin{itemize}
	\item We derive the harvested energy and uplink SE when the channels are estimated using a linear minimum mean-squared error (LMMSE) estimator for practical Rician fading channels with unknown phase shifts. We derive the SE expressions for the multi-antenna APs that are generalizations of the SE for single-antenna APs in \cite{ozge_cell_free}.
	\item We formulate the max-min fair joint AP and UE power control and large-scale fading decoding (LSFD) design problem under the harvested and transmitted power constraints at the APs and UEs.
	\item We propose an alternating optimization algorithm to achieve a solution to the proposed problem. The simulation results show that the cell-free structure and the solution found by this algorithm improve the minimum guaranteed SE of the network compared to the co-located massive MIMO and simpler power control schemes. 
\end{itemize}  
\vspace{-0.2cm}
\section{System Model}
\vspace{-0.1cm}
We consider a cell-free massive MIMO system where $L$ multiple-antenna APs are geographically distributed over a large area to serve $K$ single-antenna users with energy harvesting capability. Each AP is equipped with $N$ antennas and connected to a central processing unit (CPU) via a perfect fronthaul link. In this paper, we assume time division duplex (TDD) operation and hence channel reciprocity holds. Let $\tau_c$ denote the total number of samples per coherence interval. Each coherence interval is divided into three phases: uplink training, downlink WPT, and uplink wireless information transfer (WIT). In the uplink training phase, all the UEs send their  pilot sequences which have length $\tau_p$ to the APs, which estimate the channels to design precoding vectors for effective energy transfer and data reception. While $\tau_d$ samples are used for downlink energy transfer, the remaining $\tau_u$ samples are used for the uplink information transfer, hence we have $\tau_p+\tau_d+\tau_u=\tau_c$. In accordance with the existing literature on cell-free massive MIMO, the channel state information of users are not shared between the APs \cite{cell_free_vs_small_cell}, \cite{making_cell_free}.

Let ${\bf g}_{kl}\in \mathbb{C}^{N}$  denote the channel between the $k^{\textrm{th}}$ user and the $l^\textrm{th}$ AP. The channels are constant in each time-frequency coherence interval. We consider spatially uncorrelated Rician fading channels with unknown phase shifts, which is the first novelty of this paper in the context of cell-free massive MIMO with multiple-antenna APs. This means each channel realization can be expressed as
\vspace{-0.1cm}
\begin{align} \label{eq:channel}
& {\bf g}_{kl}=e^{j\theta_{kl}}{\bf \bar{g}}_{kl}+\tilde{\bf g}_{kl},
\end{align}
where $e^{j\theta_{kl}}{\bf \bar{g}}_{kl}\in \mathbb{C}^{N}$  denotes the line-of-sight (LOS) component. The other term of the channel, i.e., $\tilde{\bf g}_{kl}$ corresponds to non-line-of-sight (NLOS) small-scale fading and $\tilde{\bf g}_{kl} \sim \mathcal{N}_{\mathbb{C}}({\bf 0}_N,\beta_{kl}{\bf I}_N)$ where $\beta_{kl}$ is the large-scale fading coefficient which accounts for  path-loss and shadowing. Note that the vectors $\{{\bf \bar{g}}_{kl}\}$ and large-scale fading coefficients $\{\beta_{kl}\}$ describe the long-term channel effects and change more slowly compared to small-scale fading characteristics. We assume that the APs have the knowledge of $\{{\bf \bar{g}}_{kl}, \beta_{kl}\}$ corresponding to the channels between them and the UEs in accordance with the massive MIMO literature \cite{emil_book}, \cite{ozge_massive}. However, we consider a more realistic scenario where the phase shifts $\{\theta_{kl}\}$ in the LOS components are unknown due to user mobility  and assume that they are uniformly distributed in the interval $[0,2\pi)$ \cite{ozge_cell_free}.
\begin{figure*}[t!]
	\begin{align}\label{eq:lemma1}
	E_k&=\mu\tau_d\sum_{l=1}^L\sum_{i=1}^Kp_{il}\text{tr}\left({\bf \hat{R}}_{il}{\bf R}_{kl}\right)+\mu\tau_d\rho_p^2\tau_p^2\sum_{l=1}^L\sum_{i \in \mathcal{P}_k}^Kp_{il}\left( 2\beta_{kl}\Re\left\{{\bf \bar{g}}_{kl}^H{\bf \Psi}_{il}^{-1}{\bf R}_{il}{\bf \bar{g}}_{kl}\text{tr}\left({\bf R}_{il}{\bf \Psi}_{il}^{-1}\right)\right\}+\beta_{kl}^2\left|\text{tr}\left({\bf R}_{il}{\bf \Psi}_{il}^{-1}\right)\right|^2\right) \tag{14}
	\end{align} 
	\vspace{-0.6cm}
	\hrulefill
\end{figure*}
\vspace{-0.2cm}
\section{Channel Estimation}
\vspace{-0.2cm}
Let $\bm{\varphi}_k \in \mathbb{C}^{\tau_p}$ denote the pilot sequence that is assigned to the $k^\textrm{th}$ user where $||\bm{\varphi}_{k}||^2=\tau_p$. In practice, the number of users is usually larger compared to the pilot sequence length, i.e., $K>\tau_p$. Hence, so-called pilot contamination occurs. 

Note that LMMSE estimator is the MMSE estimator when the phase shifts of LOS components are known at the APs. However, deriving the MMSE estimator is non-trivial in the unknown phase shift scenario since we do not have a linear Gaussian signal model. In this paper, we will restrict ourselves to the LMMSE estimator as in \cite{ozge_cell_free}, which is the conventional benchmark in the massive MIMO literature. To obtain the LMMSE channel estimator in a simple form, let us assume that the pilot sequences are either identical or mutually orthogonal and call $\mathcal{P}_k$ the subset of users which are assigned the same pilot sequence as the $k^\textrm{th}$ user, including itself.  Then, the received pilot signal ${\bf Z}_l\in \mathbb{C}^{N\times \tau_p}$ at the $l^\textrm{th}$ AP is given by
\vspace{-0.1cm}
\begin{align}\label{eq:pilot}
& {\bf Z}_{l}=\sqrt{\rho_p}\sum_{k=1}^{K}{\bf g}_{kl}\bm{\varphi}_{k}^T+{\bf N}_l,
\end{align}
where $\rho_p$ is the transmit power of each pilot symbol and the additive noise matrix ${\bf N}_{l}\in \mathbb{C}^{N \times \tau_p}$ has i.i.d. $\mathcal{N}_{\mathbb{C}}(0,\sigma^2)$ random variables. Then, a sufficient statistics for the estimation of the $k^\textrm{th}$ user's channel is
\vspace{-0.1cm}  
\begin{align}\label{eq:suff-stats} 
& {\bf z}_{kl}=\frac{{\bf Z}_l\bm{\varphi}_k^{*}}{\sqrt{\tau_p}}=\sqrt{\rho_p\tau_p}\sum_{i \in \mathcal{P}_k}{\bf g}_{il}+{\bf n}_{kl},
\end{align}
where ${\bf n}_{kl}\triangleq{\bf N}_l\bm{\varphi}_k^{*}/\sqrt{\tau_p}\sim \mathcal{N}_{\mathbb{C}}({\bf 0}_N,\sigma^2{\bf I}_N)$. Note that ${\bf n}_{il}$ is independent of ${\bf n}_{kl}$ for $\forall i \notin \mathcal{P}_k$. Then, the phase-unaware LMMSE estimate  of ${\bf g}_{kl}$ based on \eqref{eq:suff-stats} is given by 
\vspace{-0.1cm}
\begin{align}\label{eq:lmmse} 
&{\bf \hat{g}}_{kl}=\sqrt{\rho_p\tau_p}{\bf R}_{kl}{\bf \Psi}_{kl}^{-1}{\bf z}_{kl},
\end{align}
where 
\vspace{-0.1cm}
\begin{align} 
&{\bf R}_{kl}=\mathbb{E}\{{\bf g}_{kl}{\bf g}_{kl}^H\}={\bf \bar{g}}_{kl}{\bf \bar{g}}_{kl}^H+\beta_{kl}{\bf I}_N,  \label{eq:Rkl}\\
&{\bf \Psi}_{kl}=\mathbb{E}\{{\bf z}_{kl}{\bf z}_{kl}^H\}=\rho_p\tau_p\sum_{i \in \mathcal{P}_k}{\bf R}_{il}+\sigma^2{\bf I}_N. \label{eq:Psi}
\end{align}
The channel estimate ${\bf \hat{g}}_{kl}$ and the estimation error ${\bf e}_{kl}={\bf g}_{kl}-\hat{\bf g}_{kl}$ are zero-mean uncorrelated random vectors with covariance matrices
\vspace{-0.1cm}
\begin{align}
&{\bf \hat{R}}_{kl}\triangleq\mathbb{E}\{{\bf \hat{g}}_{kl}{\bf \hat{g}}_{kl}^H\}=\rho_p\tau_p{\bf {R}}_{kl}{\bf {\Psi}}_{kl}^{-1}{\bf {R}}_{kl}, \label{eq:Rhat} \\
&{\bf C}_{kl}\triangleq\mathbb{E}\{{\bf e}_{kl}{\bf e}_{kl}^H\}={\bf {R}}_{kl}-\rho_p\tau_p{\bf {R}}_{kl}{\bf {\Psi}}_{kl}^{-1}{\bf {R}}_{kl}.
\end{align}
Note that neither channel estimate nor estimation error is Gaussian. As a result, although they are uncorrelated, they are not independent.

\vspace{-0.1cm}
\section{Downlink Wireless Power Transfer}
\vspace{-0.02cm}
In the WPT phase, each AP transmits energy to the users by using the estimated channels for maximum ratio (MR) precoding. In this paper, we will analyze  non-coherent energy transmission which do not require any synchronization among APs since it allows each AP to transmit their choice of energy symbols. During non-coherent energy harvesting phase, the signal transmitted by the $l^\textrm{th}$ AP using MR is
\begin{align} \label{eq:tr_energy}
&{\bf x}_l^E=\sum_{k=1}^K\sqrt{p_{kl}}{\bf \hat{g}}_{kl}^{*}s_{kl},
\end{align}
where $s_{kl}$  is the zero-mean unit-variance energy signal from the $l^{\textrm{th}}$ AP to the $k^{\textrm{th}}$ user. All energy signals are assumed to be independent for the ease of analysis. $p_{kl}$ is the power control coefficient of the $l^{\textrm{th}}$ AP corresponding to the $k^{\textrm{th}}$ user. The transmission power for each AP should satisfy the maximum power limit which is $\rho_d$ in the long-term, i.e.,
\begin{align} \label{eq:max_power}
&P_{l}^E\triangleq\mathbb{E}\left\{\left\Vert{\bf x}_l^E\right\Vert^2\right\}\leq \rho_d.
\end{align}
The average transmitted power $P_l^E$ for the $l^{\textrm{th}}$ AP is
\begin{align} \label{eq:Pl}
P_{l}^E=&\mathbb{E}\left\{\left\Vert\sum_{i=1}^K\sqrt{p_{il}}{\bf \hat{g}}_{il}^{*}s_{il}\right\Vert^2\right\} \nonumber \\
\stackrel{(a)}=&\sum_{k=1}^Kp_{kl}\mathbb{E}\left\{\left\Vert{\bf \hat{g}}_{kl}\right\Vert^2\right\}\stackrel{(b)}=\sum_{k=1}^Kp_{kl}\text{tr}\left({\bf \hat{R}}_{kl}\right), 
\end{align}
where we used the independence of $\{s_{kl}\}$ in $(a)$ and used \eqref{eq:Rhat} in $(b)$.
The received signal in the energy harvesting phase at the $k^{\textrm{th}}$ user is given by
\begin{align}\label{eq:rek}
r_k^E&=\sum_{l=1}^L{\bf g}_{kl}^T{\bf x}_l^E+n_k^E=\sum_{l=1}^L\sum_{i=1}^K\sqrt{p_{il}}{\bf \hat{g}}_{il}^H{\bf g}_{kl}s_{il}+n_k^E,
\end{align}
where $n_k^E \sim \mathcal{N}_{\mathbb{C}}(0,\sigma^2)$ is the additive noise at the $k^{\textrm{th}}$ user. Since the noise floor is too low for energy harvesting, we simply neglect the effect of $n_k^E$ in the average harvested energy expression in accordance with the existing literature \cite{wet_heath,swipt_cell_free,wpt1}. Then, the average harvested energy at the $k^\textrm{th}$ user during the $\tau_d$ channel uses is
\begin{align} \label{eq:Ek}
E_{k}=&\mu\tau_d\mathbb{E}\left\{\left|\sum_{l=1}^L\sum_{i=1}^K\sqrt{p_{il}}{\bf \hat{g}}_{il}^H{\bf g}_{kl}s_{il}\right|^2\right\},
\end{align}
where $\mu\in [0,1]$ is the energy harvesting efficiency of the rectifier circuit \cite{swipt_downlink,swipt_cell_free, wpt1}.
Lemma 1 presents the average harvested energy in \eqref{eq:Ek} analytically.

{\bf Lemma 1:}  If the phase-unaware LMMSE channel estimator in \eqref{eq:lmmse} is used, the average harvested energy for non-coherent downlink transmission is given in \eqref{eq:lemma1} at the top of this page.
\setcounter{equation}{14}

\begin{IEEEproof}
	The proof follows from standard expectations and omitted due to space limitation.
\end{IEEEproof}
Note that all the terms in \eqref{eq:lemma1} are positive. All users' intended signals from all the APs make a contribution to the harvested energy of each user, which is affected by the power control coefficients $\{p_{il}\}$. In addition to the first summation, having pilot contaminated channel estimates brings some additional energy terms into the second summation while it also reduces the channel estimation quality. Hence, it is not easy to quantify the effect of pilot contamination directly from the expression. As expected, with the increase in the large-scale fading coefficients $\{\beta_{kl}\}$ and the norm of the LOS parts of the channels $\{\Vert\bar{\bf g}_{kl}\Vert\}$, the harvested energy increases. Furthermore, the harvested energy is linearly proportional to the number of downlink energy symbols, $\tau_d$. However, increasing $\tau_d$ will increase the SE up to some extent since for a fixed coherence block length, $\tau_c$, an increase in $\tau_d$ necessitates a decrease in $\tau_u$ that is linearly proportional to the SE of each user as we consider in the next section.
\vspace{-0.2cm}
\section{Uplink Wireless Information Transfer}
\vspace{-0.1cm}
In the uplink information transmission phase, all the $K$ users simultaneously send their data signals to the APs. Let $q_k$ denote the symbol of the $k^{\textrm{th}}$ user, which is zero-mean with $\mathbb{E}\{|q_k|^2\}=1$, and $\eta_k\geq0$ is the corresponding transmission power. The received signal at the $l^{\textrm{th}}$ AP is given by
\begin{align} \label{eq:received_AP}
{\bf r}_l^I=\sum_{k=1}^K\sqrt{\eta_k}{\bf g}_{kl}q_k+{\bf n}_l^I, \ \ l=1,\ldots,L,
\end{align}
where ${\bf n}_l^I \sim \mathcal{N}_{\mathbb{C}}\left({\bf 0}_N,\sigma^2{\bf I}_N\right)$ is the additive white Gaussian noise. Each AP applies MR decoding for each user's information symbol before sending it to the CPU. Hence, $\tilde{r}_{kl}={\bf \hat{g}}_{kl}^H{\bf r}_l^I$ is the locally decoded signal for the $k^{\textrm{th}}$ user at the $l^{\textrm{th}}$ AP.

Then, the CPU computes a weighted sum of the locally decoded signals using the large-scale fading decoding (LSFD) method \cite{nayebi}:
\begin{align} \label{eq:LSFP_CPU}
\hat{q}_k=\sum_{l=1}^L a_{kl}^*\tilde{r}_{kl}, 
\end{align}
for the detection of the $k^{\textrm{th}}$ user's data signal where $\{a_{kl}^*\}$ are the LSFD weights. We assume that the CPU uses only the statistical knowledge of the channels in accordance with the cell-free massive MIMO literature \cite{nayebi,ozge_cell_free,making_cell_free}. Using the SE analysis technique in \cite{erik_book}, we can express the received signal at the CPU for the $k^{\textrm{th}}$ user data detection as
\begin{align}\label{eq:received_CPU2}
\hat{q}_k=\text{DS}_kq_k+\text{BU}_kq_k+\sum_{k^{\prime}\neq k}\text{UI}_{kk^{\prime}}q_{k^{\prime}}+\tilde{n}_k,
\end{align}
where $\text{DS}_k$, $\text{BU}_k$, $\text{UI}_{kk^{\prime}}$ denote the strengths of the desired signal (DS), beamforming gain uncertainty (BU), and the interference of the ${k^{\prime}}^{\textrm{th}}$ user on the $k^{\textrm{th}}$ user, while $\tilde{n}_k$ is the total noise at the CPU. $\text{DS}_k$,  $\text{BU}_k$, $\text{UI}_{kk^{\prime}}$, and $\tilde{n}_k$ are given by
\begin{align}
& \text{DS}_k=\sqrt{\eta_k}\sum_{l=1}^La_{kl}^*\mathbb{E}\left\{ {\bf \hat{g}}_{kl}^H{\bf g}_{kl} \right\}, \label{eq:DS} \\
& \text{BU}_k=\sqrt{\eta_k}\sum_{l=1}^La_{kl}^*\left({\bf \hat{g}}_{kl}^H{\bf g}_{kl}-\mathbb{E}\left\{ {\bf \hat{g}}_{kl}^H{\bf g}_{kl} \right\}\right), \label{eq:BU}
\end{align}
\begin{align}
& \text{UI}_{kk^{\prime}}=\sqrt{\eta_{k^{\prime}}}\sum_{l=1}^La_{kl}^*{\bf \hat{g}}_{kl}^H{\bf g}_{k^{\prime}l}, \ \ \ \ \tilde{n}_k=\sum_{l=1}^La_{kl}^*{\bf \hat{g}}_{kl}^H{\bf n}_l^I. \label{eq:nk}
\end{align}
Let us define the following vectors and matrices for ease of notation:
\begin{align}
& \bm{a}_k \triangleq[ \ a_{k1} \ \ldots \ a_{kL} \ ]^T\in \mathbb{C}^{L}, \label{ak} \\
& \bm{b}_k \triangleq [ \ b_{k1} \ \ldots \ b_{kL} \ ]^T\in \mathbb{C}^{L}, \ \ b_{kl}\triangleq \mathbb{E}\left\{ {\bf \hat{g}}_{kl}^H{\bf g}_{kl}\right\} \label{bk} \\
&\bm{C}_{kk^{\prime}}\in \mathbb{C}^{L \times L}, \ \ c_{kk^{\prime}}^{ll^{\prime}}\triangleq\mathbb{E}\left\{ {\bf \hat{g}}_{kl}^H{\bf g}_{k^{\prime}l}{\bf g}_{k^{\prime}l^{\prime}}^H{\bf \hat{g}}_{kl^{\prime}}\right\}, \label{Ckk}  \\
&\bm{D}_{k}\in \mathbb{C}^{L \times L}, \ \ d_{kl}\triangleq\mathbb{E}\left\{ {\bf \hat{g}}_{kl}^H{\bf n}_l^I\left({\bf n}_l^I\right)^H{\bf \hat{g}}_{kl}\right\}, \label{dk} 
\end{align}
where $c_{kk^{\prime}}^{ll^{\prime}}$ is the $(l,l^{\prime})$th element of the matrix $\bm{C}_{kk^{\prime}}$. $\bm{D}_k$ is a diagonal matrix with the $l^{\textrm{th}}$ diagonal element being $d_{kl}$.

In the following lemma we present the uplink SE which is second novelty of this paper in the context of multiple antenna cell-free massive MIMO with unknown phase-shifted Rician fading and LSFD.

{\bf Lemma 2:} The uplink  SE for the $k^{\textrm{th}}$ user with MR decoding for any finite value of $M,K,$ and $N$ is given by
\begin{align}
&R_k=\frac{\tau_u}{\tau_c}\log_2 \left(1+\text{SINR}_k\right), 
\end{align}
where the effective signal-to-noise-plus-ratio $\text{SINR}_k$ is
\begin{align}
\frac{\eta_k\left|\bm{a}_k^H\bm{b}_k\right|^2}{\bm{a}_k^H\left(\sum_{k^{\prime}=1}^K\eta_{k^{\prime}}\bm{C}_{kk^{\prime}}\right)\bm{a}_k-\eta_k\left|\bm{a}_k^H\bm{b}_k\right|^2+\bm{a}_k^H\bm{D}_k\bm{a}_k} \label{eq:Rk},
\end{align}
where the elements of $\bm{b}_k$, $\bm{C}_{kk^{\prime}}$, and $\bm{D}_k$ are given as
\begin{align}
& b_{kl}=\rho_p\tau_p{\bf \bar{g}}_{kl}^H{\bf \Psi}_{kl}^{-1}{\bf R}_{kl}{\bf \bar{g}}_{kl}+\rho_p\tau_p\beta_{kl}\text{tr}\left({\bf \Psi}_{kl}^{-1}{\bf R}_{kl}\right) , \label{eq:bkl} \\
&c_{kk^{\prime}}^{ll}=\text{tr}\left({\bf \hat{R}}_{kl}{\bf R}_{k^{\prime}l}\right)\nonumber \\
&+\mathcal{I}_{k^{\prime}\in \mathcal{P}_k}\rho_p^2\tau_p^2\Bigg(2\beta_{k^{\prime}l}\Re\left\{{\bf \bar{g}}_{k^{\prime}l}^H{\bf \Psi}_{kl}^{-1}{\bf R}_{kl}{\bf \bar{g}}_{k^{\prime}l}\text{tr}\left({\bf R}_{kl}{\bf \Psi}_{kl}^{-1}\right)\right\}\nonumber\\
&\hspace{2.3cm}+\beta_{k^{\prime}l}^2\left|\text{tr}\left({\bf R}_{kl}{\bf \Psi}_{kl}^{-1}\right)\right|^2\Bigg), \\
&c_{kk^{\prime}}^{ll^{\prime}}=\mathcal{I}_{k^{\prime}\in \mathcal{P}_k}\rho_p^2\tau_p^2\times\nonumber \\
&\Bigg(\left({\bf \bar{g}}_{k^{\prime}l}^H{\bf \Psi}_{kl}^{-1}{\bf R}_{kl}{\bf \bar{g}}_{k^{\prime}l}+\beta_{k^{\prime}l}\text{tr}\left({\bf \Psi}_{kl}^{-1}{\bf R}_{kl}\right)\right)\times\nonumber \\
&\left({\bf \bar{g}}_{k^{\prime}l^{\prime}}^H{\bf R}_{kl^{\prime}}{\bf \Psi}_{kl^{\prime}}^{-1}{\bf \bar{g}}_{k^{\prime}l^{\prime}}+\beta_{k^{\prime}l^{\prime}}\text{tr}\left({\bf R}_{kl^{\prime}}{\bf \Psi}_{kl^{\prime}}^{-1}\right)\right) \Bigg), \ \ \ l^{\prime}\neq l \\
& d_{kl}=\sigma^2\text{tr}\left({\bf \hat{R}}_{kl}\right), \label{eq:dk}
\end{align}
where $\mathcal{I}_{(.)}$ is the indicator function, i.e., $\mathcal{I}_{k^{\prime}\in\mathcal{P}_k}$ is equal to one if $k^{\prime} \in \mathcal{P}_{k}$, otherwise it is equal to zero.

\begin{IEEEproof}
	The proof follows from the fact that all the terms in \eqref{eq:received_CPU2} are uncorrelated and standard properties of circularly symmetric Gaussian random variables and uniformly distributed phase shifts. The details are omitted.
\end{IEEEproof}

We note that the effective SINR for each user is improved with an increase in the norm of its corresponding LOS components and large-scale fading of the channels to all the APs. However, an increase in these parameters lead also interference to the other UEs. Other interference sources are uncertainty in the channel estimation and pilot contamination. In fact, the signals of the other UEs that share the same pilots with the considered UE bring additional positive terms into the denominator of the effective SINR. By an intelligent power control, it is possible to maximize the minimum guaranteed SE to each UE as we consider in the next section.

\section{Max-Min Fair Joint LSFD and Power Control}
We want to maximize the minimum SE among the users by adjusting both the downlink WPT, the uplink powers, and the LSFD weights. The transmission power of the $l^{\textrm{th}}$ AP during downlink WPT phase, $P_{l}^E$ in \eqref{eq:Pl} cannot exceed the long-term maximum power limit $\rho_d$ as in \eqref{eq:max_power}. Furthermore, we require that the $k^{\textrm{th}}$ user's uplink data plus pilot energy, $\tau_u\eta_k+\tau_p\rho_p$ is upper bounded by the harvested energy $E_k$ in \eqref{eq:lemma1}. Then, the max-min fairness SE optimization problem can be cast as
\vspace{-0.4cm}
\begin{align}
& \underset{\left\{\bm{a}_k,\eta_k,p_{kl}\right\},t}{\text{maximize}} \ \ \ t \label{eq:objective} \\
& \text{subject to} \ \ \text{SINR}_k\left(\bm{a}_k,\left\{\eta_i\right\}\right)\geq t, \ \ \ k=1, \ldots,K, \label{eq:constraint1} \\
&\hspace{1.5cm}P_l^E\left(\left\{p_{il}\right\}\right)\leq \rho_d, \ \ \ l=1,\ldots,L, \label{eq:constraint2} \\
&\hspace{1.5cm}\tau_u\eta_k+\tau_p\rho_p\leq E_k\left(\left\{p_{il^{\prime}}\right\}\right), \ \ \ k=1, \ldots, K, \label{eq:constraint3} \\
&\hspace{1.5cm}p_{kl}\geq 0, \ \  l=1,\ldots,L, \ \ \eta_k\geq 0, \ \  k=1, \ldots, K, \label{eq:constraint4}
\end{align} 
where  $\text{SINR}_k$ is from \eqref{eq:Rk} and $t$ is the SINR that all users achieve. Note that the problem above is neither convex nor manageable in terms of finding the global optimum solution due to the highly coupled variables. However, an alternating optimization approach can be used in an efficient manner. The motivation for the alternating approach is explained as follows. Note that $P_l^E$ and $E_k$ are affine functions of $\{p_{il^{\prime}}\}$. Similarly, the numerator and denominator of $\text{SINR}_k$ are linear in $\{\eta_i\}$,  given the LSFD vectors $\bm{a}_k$, for $k=1,\ldots,K$. Hence, for some given $\bm{a}_k$, the optimization problem can be shown to be quasi-linear and its global optimum solution can be found using bisection search over $t$ by solving a series of simple linear programming problems \cite{linear}.  Furthermore, the LSFD vector $\bm{a}_k$ only affects the SINR of the $k^{\textrm{th}}$ user and can be found in closed form for the given uplink power coefficients $\{\eta_i\}$ by maximizing a generalized Rayleigh quotient \cite{ozge_cell_free}. Using these observations, we propose the alternating optimization algorithm which combines the closed-form LSFD vectors with the bisection search over minimum SINR as follows:
\vspace{-0.25cm}
\begin{center}
	\linethickness{0.45mm}
	\line(1,0){250}
\end{center}
\vspace{-0.2cm}
{\bf Algorithm 1:} Alternating Optimization for Max-Min Fair LSFD and Power Control 
\vspace{-0.5cm}
\begin{center}
	\linethickness{0.15mm}
	\line(1,0){250}
\end{center}
\vspace{-0.2cm}
{\bf 1) Initialization:} Choose the initial lower and upper bound for max-min SINR as $t_{\text{min}}=0$ and $t_{\text{max}}$ that is a proper positive number, respectively. Initialize $\bm{a}_k$ as all ones vector for $k=1,\ldots,K$.   \\
{\bf 2)} Set $t=\frac{t_{\text{min}}+t_{\text{max}}}{2}$. Solve the linear feasibility problem obtained by taking $\{\bm{a}_k\}$ and $t$ as constant in (\ref{eq:objective})-(\ref{eq:constraint4}), for $\{p_{kl}, \eta_k \}$. \\
{\bf 3)} If the problem in Step 2 is feasible, set the power control coefficients as the solution of this problem. Then, obtain the optimum $\{\bm{a}_k\}$ by maximizing each user's SINR as a generalized Rayleigh quotient and set $t_{\text{min}}=t^{\star}$ and $t_{\text{max}}=2t^{\star}$ where $t^{\star}$ is the minimum of the SINRs after applying LSFD. If the problem is not feasible,  set  $t_{\text{max}}=t$. \\
{\bf 4) Stop} if $t_{\text{max}}-t_{\text{min}}<\epsilon$ where $\epsilon>0$ is the tolerance parameter. Otherwise, continue with Step 2.
\vspace{-0.45cm}
\begin{center}
	\linethickness{0.15mm}
	\line(1,0){250}
\end{center}
The initial value of $t_{\text{max}}$ in Algorithm 1 can be taken as an upper bound of $t$ for the problem (\ref{eq:objective})-(\ref{eq:constraint4}). A simple upper bound can be obtained by supposing that there is only one user in the setup and maximizing the SINR of that user. If we focus on the $k^{\textrm{th}}$ user, the harvested energy, $E_k$ in \eqref{eq:Ek} is maximized by setting $p_{kl}=\rho_d/\text{tr}\left({\bf \hat{R}}_{kl}\right)$ and $p_{il}=0$, $\forall i\neq k$  by (\ref{eq:max_power})-(\ref{eq:Pl}). Let $E_k^{\star}$ denote the value of harvested energy for this setting. To maximize the $\text{SINR}_k$, we equate the total uplink transmission energy for the $k^{\textrm{th}}$ user to the harvested energy $E_k^{\star}$ in \eqref{eq:constraint3} and obtain the data power control coefficient as $\eta_k^{\star}$. We set all other power control coefficients to zero, i.e., $\eta_i=0$, $\forall i \neq k$. After maximizing the obtained generalized Rayleigh quotient for the $k^{\textrm{th}}$ user, we obtain $\text{SINR}_k^{\star}$. If we repeat this procedure for each user, we can obtain a proper upper bound for the initialization of Algorithm 1 as follows:
\vspace{-0.1cm}
\begin{align}
t_{\text{max}}=\min_{k}\text{SINR}_k^{\star}.
\end{align}
Note that in Step 3 of Algorithm 1, we change $t_{\text{max}}$ to $2t^{\star}$ if the problem in Step 2 is feasible. The reason for this update is that after LSFD, it may be possible to obtain feasible solution with $t$ larger than the $t_{max}$ that is set at the previous infeasible iterations. Note that the objective function of the problem (\ref{eq:objective})-(\ref{eq:constraint4}) is upper bounded as shown above and an improved solution is obtained at each iteration. Hence, Algorithm 1 converges.
\begin{figure}[t!]
	\includegraphics[trim={1.2cm 0cm 2.2cm 0cm},clip,width=3.57in]{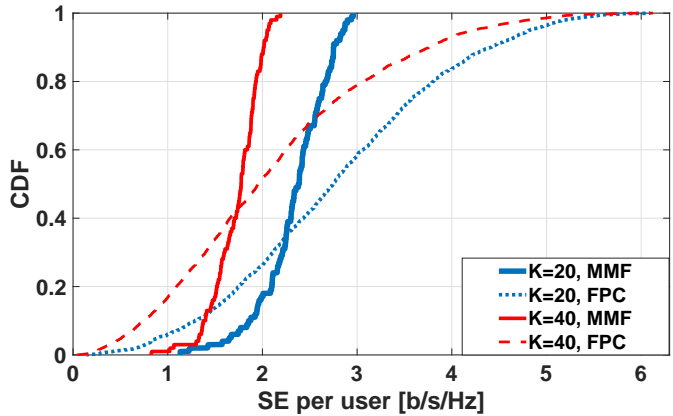}
	\vspace{-0.7cm}
	\caption{CDF of the SE per user for $L=16$ and $N=25$.}
	\label{fig:fig1a}
	\vspace{-0.4cm}
\end{figure}
\begin{figure}[t!]
	\includegraphics[trim={1.2cm 0cm 2.2cm 0cm},clip,width=3.57in]{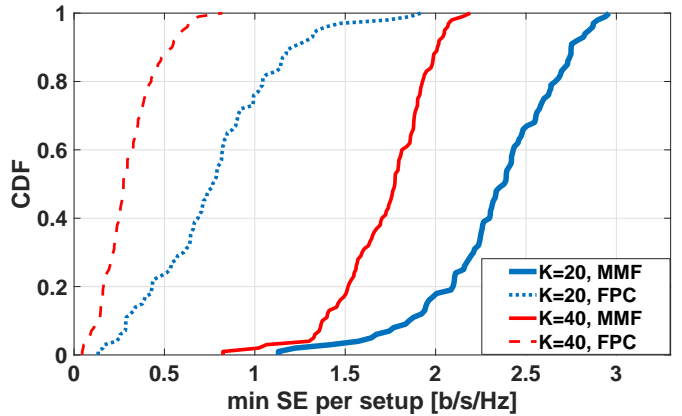}
	\vspace{-0.7cm}
	\caption{CDF of the minimum SE per setup for $L=16$ and $N=25$.}
	\label{fig:fig1b}
	\vspace{-0.4cm}
\end{figure}

\section{Numerical Results}
We will quantify the SE for different setups and compare the performance of the max-min fairness optimization with a simpler power control that is inspired by the fractional power control (FPC) scheme for the downlink information transmission in \cite{giovanni}. The 3GPP indoor hotspot (InH) model in \cite{3gpp} is used with a 3.4 GHz carrier frequency and 20\,MHz bandwidth. The large-scale fading coefficients, shadowing parameters, probability of LOS, and the Rician factors are simulated based on \cite[Table B.1.2.1-1, B.1.2.1-2, B.1.2.2.1-4]{3gpp}. The APs are uniformly distributed in a 100\,m$\times$100\,m square. For each setup, the UEs are randomly dropped and a 4\,m height difference between APs and UEs is taken into account in calculating distance. The noise variance is $\sigma^2=-96$ dBm. The uplink pilot transmission power is $-40$\,dBm. The total number of samples per coherence interval is $\tau_c=200$ with $\tau_p=5$, $\tau_d=25$, and $\tau_u=170$. The energy harvesting efficiency of the rectifier circuit, $\mu$, is 0.5. For each scenario, 100 random setups are considered.

In the first scenario, we consider $L=16$ APs, each with $N=25$ antennas. The maximum power of each AP is $\rho_d=250$\,mW. MMF stands for the proposed max-min fairness optimization. For the other scheme that is  shown by FPC, the power control coefficient $p_{kl}$ is proportional to $1/\sqrt{\text{tr}({\bf \hat{R}}_{lk})}$ and they are scaled such that total transmission power is $\rho_d$ for each AP in accordance with the power control scheme \cite{giovanni}. Each UE's power control coefficient $\eta_k$ is adjusted such that total uplink transmission energy is equal to the harvested energy in the downlink.

In Fig.~1, we plot the cumulative distribution function (CDF) of the individual SE per user. We notice that the 90\% and 95\% likely SE (i.e., where
the CDF is 0.1 and 0.05, respectively) is better for the proposed max-min fair design by 43\% and 84\% in comparison to the FPC for $K=20$ UEs. The relative improvement by the max-min fairness is larger for $K=40$ UEs, i.e., by 84\% and 159\%, for 90\% and 95\% likely SE, respectively. 

In order to see the fairness improvement of the proposed algorithm, we plot the CDF of the minimum SE of all the UEs per setup in Fig.~2 for the same scenario. For both $K=20$ and $K=40$, the minimum SE of the network improves substantially and larger SE is guaranteed for all the UEs.

\begin{figure}[t!]
	\includegraphics[trim={1.2cm 0cm 2.2cm 0cm},clip,width=3.57in]{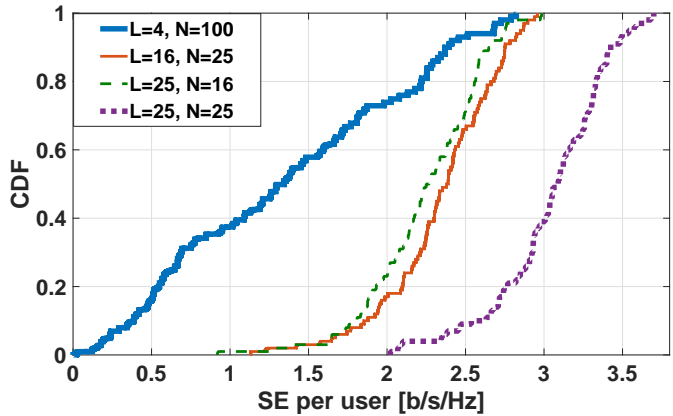}
		\vspace{-0.7cm}
	\caption{CDF of the SE per user for $K=20$.}
	\label{fig:fig2}
			\vspace{-0.4cm}
\end{figure}

In Fig.~3, we quantify the impact of the number of APs, $L$, and antennas per AP, $N$, for $K=20$ UEs. The maximum transmission power for each AP is $\rho_d=4/L$\,W. Hence, maximum total transmit power for the whole AP network is $4$\,W for a fair comparison. Note that the SE for all the UEs with co-located massive MIMO with $L=1$ and $N=400$ antennas is relatively very small and not included in the Fig.~3. For the first three lines in Fig.~3, the number of total antennas throughout all the area is $LN=400$. We notice that the 90\% likely SE is improved by 379\% by increasing the number of APs from $L=4$ to $L=16$. However, there is a slight performance decrease when we increase it to $L=25$ by keeping the total number of antennas the same. We believe that this is due to the increased number of local power constraints in \eqref{eq:constraint2}, which prevent more improvement. However, if we increase the number of antennas per AP to $N=25$, we now see the positive impact of jointly increasing the number of APs and total number of antennas, $LN$, where each UE's SE is significantly improved.

In the above simulations, the duration of pilot, energy and data transmission are fixed. In the journal extension of this paper \cite{cell_free_wpt}, it is shown that changing the energy duration, $\tau_d$ has a less significant effect on the SE compared to the other system parameters and channel estimation quality.

\section{Conclusion}
In this paper, we have derived the uplink SE of the wireless-powered cell-free massive MIMO in Rician fading with LOS components that are phase-shifted in each coherence block. The UEs harvest energy from the RF signal that APs direct to them by MR processing in the downlink. Then, they use some portion of the harvested energy for the uplink data transmission. We optimize both the downlink WPT and uplink WIT power control coefficients together with LSFD weights at the CPU to maximize the minimum guaranteed SE for all the UEs. An alternating optimization algorithm is proposed for solving the non-convex problem. Simulation results show the fairness improvement of the proposed algorithm compared to another state-of-the-art power control scheme that was originally proposed for downlink information transmission. Furthermore, increasing the number of APs to a certain extent improves the SE.

\end{document}